\documentclass[aps,prb,twocolumn,showpacs]{revtex4-1}
\usepackage{graphics}
\usepackage{subfigure}
\usepackage{epsfig}
\usepackage{epsf,epic}
\usepackage{color}
\usepackage{marvosym}
\usepackage{subfigure}
\usepackage{amsmath}
\usepackage{amssymb}
\usepackage{amsfonts}
\usepackage{wrapfig}
\usepackage{pstricks}
\usepackage{multirow}
\usepackage{bm}
\usepackage{dcolumn}
\newcommand{\etal}{\textit{et al.\ }}

\begin{document}
\title{Computational Study of Electron Paramagnetic Resonance Spectra for Li and Ga Vacancies in LiGaO$_2$}

\author{Dmitry Skachkov}
\email{Corresponding author dmitry.skachkov@case.edu, Current address: Department of Physics, University of Florida, Gainesville, Florida 32611}
\author{Walter R. L. Lambrecht}
\affiliation{Department of Physics, Case Western Reserve University, 10900 Euclid Avenue, Cleveland, Ohio 44106-7079, USA}
\author{Klichchupong Dabsamut}
\author{Adisak Boonchun} 
\affiliation{Department of Physics, Faculty of Science, Kasetsart University, Bangkok 10900 Thailand}

\begin{abstract}
  A  first-principles computational study of the Electron Paramagnetic  Resonance (EPR) parameters of Li and Ga vacancies in LiGaO$_2$ is presented. In the EPR active charge states  $V_\mathrm{Li}^0$
  and $V_\mathrm{Ga}^{2-}$, the spin is localized on one of the O neighbors
  of the vacancy. We compare the calculated EPR parameters for
  spin density localized on different O neighbors. Good agreement with
  experiment is obtained for both the $g$-tensor values and
  principal axes orientations and the superfhyperfine interaction parameters
  supporting the prior experimental
  identification of which O the spin is localized on.  
  The $g$-tensor orientations are found to be close to the bond
  rather than the crystalline axes. The high energy of formation
  of $V_\mathrm{Ga}$  compared to $V_\mathrm{Li}$ also explains
  why $V_\mathrm{Ga}$ were only observed after high energy irradiation
  while $V_\mathrm{Li}$ were found in as grown samples. On the other hand,
  the transition levels and Fermi level position explain why $V_\mathrm{Li}$
  required ionization from the $-1$ to $0$ charge state to become active
  while $V_\mathrm{Ga}$ were already found in the $q=-2$ EPR active state. 
\end{abstract}

\maketitle
LiGaO$_2$ is an ultra-wide-band-gap material with a wurtzite-like crystal structure\cite{Marezio65,Ishii98} and experimental band gap of $\sim$5.3--5.6 eV
at room temperature.\cite{Wolan98,Johnson2011,Chen14,Ohkubo2002} 
It can be grown in bulk form by
the Czochralsky method\cite{Marezio65} and has been suggested as a useful
substrate for GaN but can also be grown by epitaxial method
on ZnO and vice versa. Mixed ZnO-LiGaO$_2$ alloys have also been reported.
\cite{Omata08,Omata11} In fact, this material can be viewed
as a I-III-VI$_2$ analog of the II-VI material ZnO by substituting the
II-element Zn by a group I (Li) and a group III (Ga) in an ordered fashion
on the wurtzite lattice. 
It has been considered for piezoelectric properties\cite{Nanamatsu72,Gupta76,Boonchun2010} in the past and is for the most part considered an
insulator. However, Boonchun  and Lambrecht \cite{Boonchun11} suggested
it might be
worthwhile considering as a semiconductor electronic material and
showed in particular that it could possibly be n-type doped by Ge.
In view of the recent interest in $\beta$-Ga$_2$O$_3$ as ultra-wide
semiconductor for power electronics, which is also n-type by
doping with Si, Sn or Ge, this makes LiGaO$_2$ worth revisiting,
in particular from the point of view of defects and doping. 

Recently, Electron Paramagnetic Resonance (EPR) experiments on irradiated 
samples of LiGaO$_2$ were reported by Lenyk \etal\cite{Lenyk18} and 
reported EPR signals for both the $V_\mathrm{Ga}$ and $V_\mathrm{Li}$.
Here we present a computational study of the EPR parameters of these defects
and in particular compare the calculated $g$-tensors and
superhyperfine (SHF) interactions
with both Ga and Li neighbors of the O on which the spin is localized
for different possible localization sites of the spin.
We will show that this confirms the experimentally deduced models
for the spin-localization.  

The $g$-tensor is calculated using the Gauge Including Projector Augmented Wave (GIPAW) method.\cite{Pickard01,Pickard02,Gerstmann10,Ceresoli10} 
This is  a Density Functional Perturbation Theory (DFPT) method
to calculate the linear magnetic response of a periodic system
onto  an external magnetic field. It is implemented in the code QE-GIPAW,\cite{gipaw} which is integrated within the Quantum Espresso package.\cite{QE-2009}
At present the QE-GIPAW  code is not yet capable of dealing with orbital
dependent density functionals such as DFT+U \cite{Anisimov91,Anisimov93,Liechtenstein95} or hybrid functionals.\cite{HSE03,HSE06}
The latter are required to ensure a strong localization of the
spin-density on a single O. We thus use DFT with on-site Coulomb corrections
$U$ on O-$p$ orbitals with the
Perdew-Burke-Ernzerhof (PBE) generalized gradient approximation (GGA)
to relax the structure
and also calculate the SHF interactions at this GGA+U level but calculate the
$g$-tensor using wavefunctions at the PBE-GGA level while keeping the
structure fixed. This procedure was found to be adequate in prior work
on EPR parameters in $\beta$-Ga$_2$O$_3$.\cite{Skachkov19,Skachkov19-mg}
The GGA+U structures were in good agreement with a previous
study of the same defects\cite{Dabsamut} using the Heyd-Scuseria-Ernzerhof 
(HSE) hybrid functional.

\begin{figure*}
  \includegraphics[width=16cm]{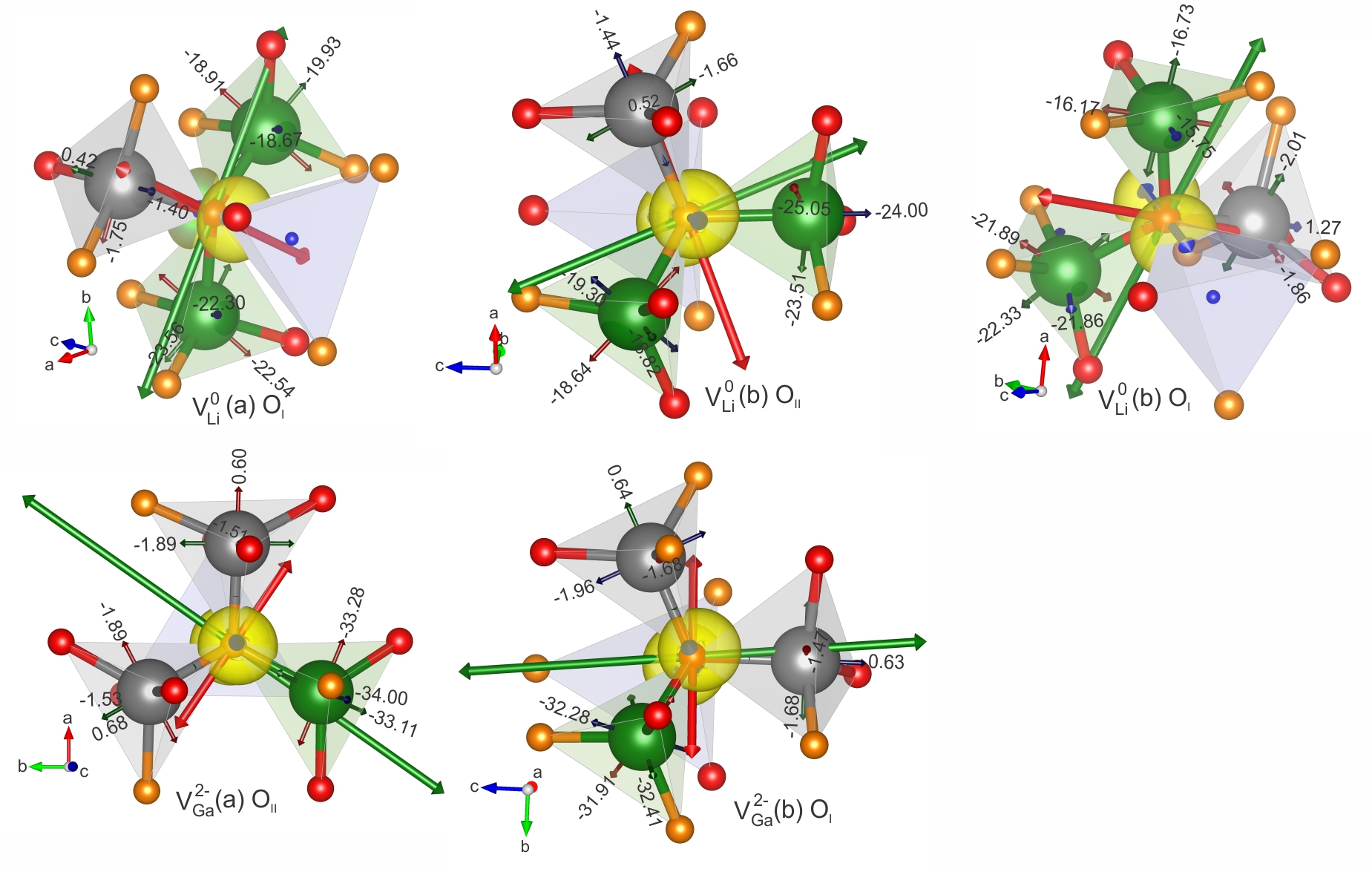}
  \caption{EPR models for various localizations of the spin density near $V_\mathrm{Li}$ (top) and $V_\mathrm{Ga}$ (bottom). 
    The spin density is shown as the yellow isosurfaces, the $g$-tensor is
    shown as large green, red, and blue double arrows with size proportional to $\Delta g$ and
    direction indicating
    the principal axes, green spheres indicate Ga with strong hyperfine
    interaction, silver spheres indicate Li, red spheres O$_I$, and orange spheres O$_{II}$. The small
    arrows on each atom indicate the principal axes and sizes of the
    hyperfine tensor ${\bf A}$. The light blue tetrahedron and blue dots indicate the vacancy site. \label{Fig1}}
\end{figure*}

We focus on the EPR active states $V_\mathrm{Ga}^{2-}$
and $V_\mathrm{Li}^{0}$ which both correspond to a $S=1/2$ single unpaired
electron state. We find that in the DFT+U approach with a value of $U=4$ eV
on O-$p$ orbitals, the spin-density becomes well localized on a single O
$p$-orbital but depending on the initial displacements given to the O,
we can get it to localize on different O neighbors. Keeping this
relaxed structure, the spin then stays localized on the single O
even when recalculating it in GGA. We need to distinguish the following
O-sites. First in the crystal structure, the O$_I$ sits on top of Li
and the O$_{II}$ sits on top of Ga in the ${\bf c}$ direction.
Secondly,  we call an O {\sl apical} if it sits right above the vacancy (in the
{\bf c}-direction) and {\sl basal} if it lies in the ${\bf ab}$-plane below it. 
The basal plane O can still be either O$_I$ or O$_{II}$. 
Our results for all the cases considered are summarized in Fig. \ref{Fig1},
the $g$-tensors are summarized in Table \ref{tabeprg} and the
SHF tensors are given in Table \ref{tabeprA}.

\begin{table}
  \caption{Calculated $g$-tensor for Li and Ga vacancies. In the calculated
    results, the $g$-tensor 
    is given in terms of three principal values followed by the $\theta$ (polar)  and $\phi$ (azimuthal) angles in degrees
    measured from ${\bf c}$ and ${\bf a}$ respectively. The experimental
    values are along the directions indicated.
           \label{tabeprg}
   }
  \begin{ruledtabular}
    \begin{tabular}{l|l|ccc}
      Model         &          &    &  $g$-tensor  &      \\ \hline
V$_\mathrm{Li}$ (a) O$_I$ &    & 2.0373 & 2.0288 & 2.0078     \\
                    & $\theta$ &  89    &    89  &   2         \\
                    & $\phi$   &  -34   &   56   &           \\ \hline
V$_\mathrm{Li}$ (b) O$_{II}$&  & 2.0302 & 2.0119 & 2.0356    \\
                    & $\theta$ &   69   &   71   &  29         \\
                    & $\phi$   &  17    &   -65  &  64         \\ \hline
V$_\mathrm{Li}$ (b) O$_{I}$&  & 2.0403 & 2.0125 & 2.0301    \\
                    & $\theta$ &   90   &   73   &  17         \\
                    & $\phi$   &  -27   &   64   &  62         \\ \hline
Expt.\cite{Lenyk18} &          & 2.0088 & 2.0205 & 2.0366 \\ 
                    &          &  $a$   &   $b$  &   $c$   \\ \hline \hline
V$_\mathrm{Ga}$ (a) O$_{II}$  &  & 2.0220 & 2.0514 & 2.0078    \\
                    & $\theta$ &  84    &    88  &   6         \\
                    & $\phi$   & -35    &    55  &             \\ \hline
V$_\mathrm{Ga}$ (b) O$_I$      &        & 2.0081 & 2.0228 & 2.0449  \\
                    & $\theta$ &  88    &    85  &   5    \\
                    & $\phi$   &  -16   &    74  &       \\ \hline
Expt.\cite{Lenyk18} &          & 2.0155 & 2.0551 & 2.0032 \\ 
                    &          &  $a$   &   $b$  &   $c$   \\ 
    \end{tabular}
  \end{ruledtabular}
\end{table}

\begin{table*}
  \caption{Calculated $A$-tensors of Li and Ga vacancies (in Gauss).
    The principal values and principal axes directions are given in spherical angles $\theta$ and $\phi$ 
    ($\theta$ is measured from ${\bf c}$ and $\phi$ from ${\bf a}$). The experimental results give the values along specific
    crystallographic directions as indicated and the two Ga neighbors are not distinguished because in principle equivalent.
    The Li hyperfine $A$ were not yet measured. 
           \label{tabeprA}
   }
  \begin{ruledtabular}
    \begin{tabular}{l|l|ccc|ccc|ccc|ccc}
      Model         &          &        &  Ga1   &        &         & Ga2   &          &       & Li1   &         &      & Li2            \\ \hline
V$_\mathrm{Li}$ (a) O$_I$ &    & -19.93 & -18.91 & -18.67 & -23.56  & -22.54& -22.30   & 0.42  & -1.75 & -1.40   &      &      &    \\
                    & $\theta$ &   80   &    61  &   31   &  80     &   61  &  31      & 81    & 90    & 9       &      &      &       \\
                    & $\phi$   &   -51  &    33  &   57   &  -51    &   33  &  57      & 62    & -28   & 60      &      &      &       \\ \hline
V$_\mathrm{Li}$ (b) O$_{II}$ & & -18.82 & -18.64 & -19.30 &  -23.51 & -25.05& -24.00   & -1.66 & 0.52  & -1.44   &      &      &     \\
                    & $\theta$ &   80   &   88   &  10    &   68    & 48    & 51       & 34    &  68   & 66      &      &      &       \\
                    & $\phi$   &  -34   &   56   &  -48   &   -85   & 27    & -14      & 62    &  9    & -70     &      &      &       \\ \hline
V$_\mathrm{Li}$ (b) O$_{I}$  & & -16.17 & -16.73 & -15.75 &  -21.89 & -21.86& -22.33   & 1.27  & -2.00 & -1.86   &      &      &     \\
                    & $\theta$ &   34   &   88   &  57    &   33    & 58    & 83       & 12    &  89   & 78      &      &      &       \\
                    & $\phi$   &   81   &  -12   &   77   &   34    & 49    & -46      & 65    &  -31  & 59      &      &      &       \\ \hline
Expt.\cite{Lenyk18} &          &  24.30 &  25.00 &  25.30 &   24.30      & 25.00      & 25.30          &       &       &  \\ 
                    &          &  $a$   &   $b$  &   $c$  &   $a$   &   $b$ &   c& \\ \hline \hline
V$_\mathrm{Ga}$ (a) O$_{II}$ & & -33.11 & -33.28 & -34.00 &         &       &          & -1.89 & 0.60  & -1.51   & 0.68 & -1.89& -1.53    \\
                    & $\theta$ &  67    &    90  &  23    &         &       &          & 89    & 80    & 10      & 80   &  88  & 10      \\
                    & $\phi$   &  67    &   -23  &  68    &         &       &          & 90    & 0     &         & -62  &  28  &       \\ \hline
V$_\mathrm{Ga}$ (b) O$_I$  &   & -32.41 & -31.91 & -32.28 &         &       &          & -1.68 & -1.47 & 0.63    & 0.64 & -1.68& -1.96  \\
                    & $\theta$ &  72    &    55  &  41    &         &       &          & 89    & 80    & 10      & 66   & 72   & 31  \\
                    & $\phi$   &  56    &   -47  &   -11  &         &       &          & 88    & -2    &         & -70  & 11   & 68  \\ \hline
Expt.\cite{Lenyk18} &          &  37.50 &  37.40 &  35.90 &        &       &          &       &       &  \\
                    &          &  $a$   &   $b$  &   $c$  &    \\ 
    \end{tabular}
  \end{ruledtabular}
\end{table*}

For the $V_\mathrm{Ga}$ we examine both the apical and basal plane O$_I$
as atom for the hole to localize on.  As shown in Fig. \ref{Fig1} (lower-left)
and in Table \ref{tabeprg}
we find the $g$-tensor for the apical O$_{II}$ has its smallest
$g$ along the direction of the spin density, which is along ${\bf c}$.
This agrees with experiment.\cite{Lenyk18} The  largest
principal axis (principal axis corresponding to largest $\Delta g$) 
in the ${\bf ab}$-plane is $55^\circ$  from ${\bf a}$ so closer to ${\bf b}$
which also agrees with experiment. In fact it is close to the O$_{II}$-Ga
direction. 
The $\Delta g$ themselves are in agreement to
about $\pm0.005$. For the basal plane O$_I$ location of the spin,
(Fig. \ref{Fig1} lower-right)
on the other hand the  main principal axis of the
$g$ tensor is along ${\bf c}$. In both cases it is perpendicular to the spin
direction of the spin density $p$-orbital which corresponds to the lowest
$\Delta g$ direction. The SHF interaction (given in Table \ref{tabeprA}) in both  cases is with
one Ga atom because obviously the O on which the spin has localized
has already lost one of its Ga neighbors and each O is coordinated with
two Ga and two Li. It is called a SHF interaction because the
nucleus with which the electron spin is interacting is not on the atom on which
the spin is localized but one of its neighbors. The hyperfine tensor $A$
is nearly isotropic with a value of about 33 G in excellent agreement with the
experimental values of about 37 G. Our values are about 10\% underestimated.
In agreement with experiment we find a slightly larger $A$ component in the
${\bf c}$ direction for the apical O. 
The hyperfine with O is not
observed because O is more than 99.9 \% isotopically in a form which does
not carry a nuclear spin. The hyperfine
principal axes are indicated by the small arrows in Fig. \ref{Fig1}
and are seen to
be close to the bond directions rather than the overall crystal axes. 
While the Li SHF interactions were not observed we give the calculated
values for them in Table \ref{tabeprA} in case future measurements would be able to measure them. The reason why they are much smaller is that the
atomic wavefunction on the Li nuclear sites are much smaller than on the Ga.

The $V_\mathrm{Li}$ with spin localized on an apical O$_I$ has its main
$\Delta g$-tensor component at about 30$^\circ$ from the ${\bf a}$-axis
and its lowest
component and spin density along ${\bf c}$ as can be seen in Fig. \ref{Fig1}
(upper left). 
This, however does not agree with
the experimental data of Lenyk \etal\cite{Lenyk18} who find
the $\Delta g$ tensor to be oriented with its highest value along ${\bf c}$.
We have calculated two distinct configurations with spin localized on
a basal plane O$_{I}$ and O$_{II}$ (See Table \ref{tabeprg}).
For the O$_{II}$ case we find that the lowest $\Delta g$ is coincident
with the direction of the spin density and is close to the bond direction
from $V_\mathrm{Li}$ to the O$_{II}$. So, it is tilted away from the ${\bf ab}$-plane
by about 30$^\circ$ and close to 60$^\circ$ degrees from the ${\bf a}$-axis.
Note, however, that there is an equivalent O$_{II}$ along the ${\bf a}$ axis
in the ${\bf ab}$-plane projection, which is simply 120$^\circ$ rotated from
the one reported in Table \ref{tabeprg}. The highest $g$-component
principal axis has to be perpendicular to this and indeed
we find it to be tilted about
30$^\circ$ from the ${\bf c}$-axis. This model agrees closely
with the one proposed by Lenyk \etal\cite{Lenyk18} with the sole difference
that they consider the equivalent O$_{II}$ in the ${\bf a}$-direction.
As the authors
mention, the occurrence of several distinct magnetic orientations prevents
them from carrying out a full study of the angular variation with
magnetic field because of the overlap of different signals.
As for the O$_{I}$ basal plane, neighbor, in that case the lowest $g$-component
is along the corresponding V$_\mathrm{Li}-\mathrm{O}_I$ direction at about 60$^\circ$ from the ${\bf a}$ axis.
However, the largest $g$ is then found at about $-30^\circ$ from
${\bf a}$ and tilted toward the basal plane. 
This does not agree with the center identified by Lenyk \etal\cite{Lenyk18}
 
The SHF splitting in $V_\mathrm{Li}$ with spin localized on O$_{II}$
is with two nonequivalent Ga. Although all Ga atoms are equivalent in the
perfect crystal, the local symmetry is broken. The Ga with smaller
SHF $A$ tensor lies closer to the $V_\mathrm{Li}$ than the other
which lies opposite to it from the $O_{II}$ on which the spin is localized.
In the experiment, also a slightly nonequivalent Ga-SHF splitting
was reported but they estimated the $A$'s to differ by only 4\%
whereas we find them to differ by about 20\%. In agreement with experiment
the $A$ tensors are found to be nearly isotropic.  The experimental
value for the SHF splitting is closer to the larger of the two
calculated $A$ and is in good agreement with experiment. 
For the apical O$_I$ case, one would expect the two Ga neighbors to
be equivalent but in the calculation, they are still found to differ by 
about 20\%, which may result from the symmetry breaking in the relaxation
calculation.

Having identified the apical O as the location of the spin density near
a $V_\mathrm{Ga}$ and the basal plane O near a $V_\mathrm{Li}$ that
best agree with experiment, we may ask whether these indeed correspond
to the lowest total energy. It turns out, however, that the energy differences
between these different localization sites is quite small.
We find that the $V_\mathrm{Ga}$ has 0.01 eV higher energy
per 128 atom cell in the apical than the basal plane
site within PBE0,\cite{PBEh}
(this is a hybrid functional with 25 \% exact and unscreened exchange)
so opposite to the experimental identification. 
For the $V_\mathrm{Li}$ it is the apical oxygen that was found to have the lower
energy by 0.002 eV. In the HSE functional, the apical site was found
upon automatic relaxation for both cases.  Clearly these energy differences
are too small to trust within DFT or at least this is very challenging
for any level of theory. 
Therefore we expect that several of these slightly different forms of the
vacancy EPR centers with spin localized on different O-neighbors
could be present in experiment but the overlap of these signals would
make it difficult to disentangle them. The
apical O$_{II}$ for $V_\mathrm{Ga}$ and basal-plane O$_{II}$ for $V_\mathrm{Li}$
agree best with the experimental observations but in the $V_\mathrm{Li}$ case,
there would still be two differently oriented forms of this same defect.

Finally, we address the question under what conditions these EPR signals
were observed. A hybrid functional study of the native defects in LiGaO$_2$
was recently presented by some of us.\cite{Dabsamut}
From that study, we find that the $V_\mathrm{Li}^0$
has lower energy than the $V_\mathrm{Ga}^0$
for all chemical conditions as restricted by the formation of competing binary
compounds and under Li-poor conditions can be lower than 1 eV.
The $V_\mathrm{Ga}$ usually has quite high energy (10 eV for Ga-rich conditions
and $\sim$4.5 eV  under the most Ga-poor, Li-rich conditions allowed) 
and is not expected to
occur in significant concentration in equilibrium. 
In contrast,  the $V_\mathrm{Li}^{-}$
is found to be the major acceptor compensating the Ga$_\mathrm{Li}^{2+}$
antisite and is thus expected to be present in the as grown samples. 
The $V_\mathrm{Li}$ occurs in 0 and  $-1$  charge states, the former of which
contains an unpaired spin and is hence EPR active. Its $0/-$ transition
level lies at 1.03 eV above the valence band maximum (VBM). 
The Ga-vacancy accommodates four charge states, 0, $-1$,$-2$, $-3$.
The Fermi level is pinned by the compensation of Ga$_\mathrm{Li}^{2+}$
antisites with $V_\mathrm{Li}^{-}$  and to some extent by 
Li$_\mathrm{Ga}^{2-}$ in Li-rich conditions. 
In both cases, the Fermi level lies deep below the conduction band between 2.7-3.8 eV
above the VBM 
straddling the $2-/3-$ transition level of the $V_\mathrm{Ga}$, which occurs at
3.3 eV. The $V_\mathrm{Ga}$,  once it is formed, may thus be expected
to be found in the EPR active $q=-2$ charge state in particular for
the deeper Fermi level position, which occurs for more realistic
assumptions of the O-chemical potential. 

The above findings agree with the observations of Lenyk \etal\cite{Lenyk18}
that high-energy particle irradiation is required to create the $V_\mathrm{Ga}$.
However, the fact that they
do not require to be optically activated once formed indicates a $2-$
charge state after irradiation. On the other hand the Li-vacancies
were found to be present already in as-grown material. This however does
not imply the material was Li-poor. Even under both Li and Ga rich
conditions, the $V_\mathrm{Li}$ has  an energy of formation
significantly lower than that of Ga. However, the fact that its $0/-$
level lies only 1.02 eV above the VBM clearly explains why the Li must
be activated optically by removing an electron from it. In the experiments
by Lenyk \etal\cite{Lenyk18} this is achieved by application X-rays.

In conclusion, our first-principles calculations confirm the experimental assignment of the
EPR centers of $V_\mathrm{Ga}$ and $V_\mathrm{Li}$ by Lenyk \etal\cite{Lenyk18}
For the $V_\mathrm{Ga}$ the spin is localized on an apical O$_{II}$
and for the $V_\mathrm{Li}$ it is on a basal plane O$_{II}$. The orientations
of the principal axes of the $g$-tensor and $A$-tensors are found to be
closely related to the bond directions and in the $V_\mathrm{Li}$ case
two different orientations of the  defect center with respect to the
crystal axis should exist with overlapping spectra.  The  EPR parameters
for alternative localizations of the spin on different O-neighbors
were also calculated and found to be different. As these different
forms of localization of the spin have total energies close to each other
they might possibly occur in the real systems and we hope that
providing the associated parameters here could assist in disentangling
these different EPR centers. For the cases observed so far, our $g$-tensor
and SHF interaction parameters are in good agreement with experiment. 
Our calculations also explain why the
$V_\mathrm{Ga}$ defects require high energy radiation to be formed
but no further optical activation while the opposite is the case for the
$V_\mathrm{Li}$.

{\bf Acknowledgements:} The work at CWRU was supported by
  the U.S. National Science Foundation under grant No. 1755479.
  This work used the Extreme Science and Engineering Discovery Environment (XSEDE) 
  Stampede2 at the UT Austin through allocation TG-DMR180118.

\bibliography{dft,ligao2,gipaw,ldau,defects}
\end{document}